\documentclass[twocolumn,showpacs,superscriptaddress,amsmath,amssymb]{revtex4}

\usepackage{float,epsfig}

\begin{document}

\title{Nuclear Structure Calculations with Low-Momentum Potentials in 
a Model Space Truncation Approach}
\author{L. Coraggio}
\affiliation{Dipartimento di Scienze Fisiche, Universit\`a
di Napoli Federico II, \\ and Istituto Nazionale di Fisica Nucleare, \\
Complesso Universitario di Monte  S. Angelo, Via Cintia - I-80126 Napoli,
Italy}
\author{A. Covello}
\affiliation{Dipartimento di Scienze Fisiche, Universit\`a
di Napoli Federico II, \\ and Istituto Nazionale di Fisica Nucleare, \\
Complesso Universitario di Monte  S. Angelo, Via Cintia - I-80126 Napoli,
Italy}
\author{A. Gargano}
\affiliation{Dipartimento di Scienze Fisiche, Universit\`a
di Napoli Federico II, \\ and Istituto Nazionale di Fisica Nucleare, \\
Complesso Universitario di Monte  S. Angelo, Via Cintia - I-80126 Napoli,
Italy}
\author{N. Itaco}
\affiliation{Dipartimento di Scienze Fisiche, Universit\`a
di Napoli Federico II, \\ and Istituto Nazionale di Fisica Nucleare, \\
Complesso Universitario di Monte  S. Angelo, Via Cintia - I-80126 Napoli,
Italy}
\author{T. T. S. Kuo}
\affiliation{Department of Physics, SUNY, Stony Brook, New York 11794}

\date{\today}

\begin{abstract}
We have calculated the ground-state energy of the doubly magic nuclei 
$^4$He, $^{16}$O and $^{40}$Ca within the framework of the Goldstone 
expansion starting from various modern nucleon-nucleon potentials. 
The short-range repulsion of these potentials has been renormalized by 
constructing a low-momentum potential $V_{\rm low-k}$. 
We have studied the connection between the cutoff monemtum $\Lambda$
and the size of the harmonic oscillator space employed in the
calculations. 
We have found  a fast convergence of the results with a limited number
of oscillator quanta.
\end{abstract}


\maketitle

\section{Introduction}
One of the well-known important features of realistic nucleon-nucleon ($NN$)
potentials is their strong repulsive behavior in the high-momentum regime.
This implies that when performing nuclear structure calculations
within a perturbative approach it is unavoidable to renormalize the
$NN$ potential.
The renormalization may be achieved by constructing a potential that 
takes into account the high momentum components of the original
potential in an effective way.
For example, the well-known Brueckner reaction matrix $G$ \cite{bethe71} 
is an energy-dependent effective interaction, defined in a model space
$P$, obtained projecting out all two-particle excitations above a
chosen Fermi surface.
The main drawback of such a procedure is that the $G$ matrix is energy 
dependent, which stems from the fact that it does not fulfill a
decoupling condition between the model space $P$ and its complement $Q$.

Recently, we have proposed \cite{bogner01,bogner02} a new method to 
renormalize the bare $NN$ interaction, which is proving to be an 
advantageous alternative to the use of the Brueckner $G$ matrix.
We define a low-momentum $P$-space up to a cutoff momentum $\Lambda$
and derive from the original $V_{NN}$ an effective low-momentum potential 
$V_{\rm low-k}$ that satisfies the decoupling condition between the
low- and high-momentum spaces.
This $V_{\rm low-k}$ is a smooth potential which preserves exactly the 
on-shell properties of the original $V_{NN}$ and is suitable for being
used directly in nuclear structure calculations. 

In applying the low-momentum $NN$ potential $V_{\rm low-k}$ to nuclear 
structure calculations, an important step is the determination of the 
decimation momentum $\Lambda$ for $V_{\rm low-k}$. 
There are two considerations: first, $\Lambda$ is based on the
separation of scale idea of the renormalization-group effective field
theory (RGEFT) approach. 
In low-energy nuclear physics, one is interested in phenomena at 
low-energy scale and consequently the details of the short-distance 
(high-energy scale) physics are unimportant.
This leads to the derivation of the low-momentum potential $V_{\rm low-k}$ 
by integrating out the high-momentum components of modern $NN$
potentials beyond a cutoff momentum $\Lambda$.

The second consideration is that all modern $NN$ potentials are
constructed to fit the empirical phase shifts up to the inelastic 
threshold $E_{\rm lab} \simeq 350$ MeV, which corresponds to a maximum 
relative momentum $\simeq 2.1 \; {\rm fm}^{-1}$.

In the past few years, we have profitably employed this technique in
realistic nuclear structure calculations for both doubly closed-shell
nuclei, within the framework of the Goldstone expansion 
\cite{coraggio03,coraggio05}, and open-shell nuclei within the
multilevel shell-model framework 
\cite{bogner01,bogner02,coraggio02a,coraggio02b,covello02}.
In all these works we have used for the cutoff momentum the value 
$\Lambda=2.1 \; {\rm fm}^{-1}$.

It is interesting, however, to investigate the relation of the cutoff 
momentum $\Lambda$ to the dimension of the configuration space in the 
coordinate representation, where actually our calculations are performed.
The study of such a relation is the aim of the present work, where we
show how the choice of a cutoff momentum implies a maximum value for
the energy of the two-nucleon system, the latter introducing a simple 
criterion to choose the two-nucleon model space.

To verify the reliability of this approach we calculate, in the
framework of the Goldstone expansion, the ground state (g.s.)
properties of doubly closed-shell nuclei starting from different $NN$ 
potentials renormalized through the $V_{\rm low-k}$ procedure.

The paper is organized as follows. 
In Sec. II we give an outline of our calculations.
A detailed discussion of the convergence properties of the
perturbative calculations is presented in Sec. III.
Sec. IV is devoted to the presentation and discussion of our
results for the three nuclei $^{4}$He, $^{16}$O, and $^{40}$Ca. 
Some concluding remarks are given in Sec. V.

\section{Outline of calculations}
As already mentioned in the Introduction, we renormalize the short-range 
repulsion of the bare $NN$ potential $V_{NN}$ by integrating out its high
momentum components through the so-called $V_{\rm low-k}$ approach. 
A detailed description of this procedure can be found in Refs. 
\cite{bogner02,bogner03}.
The renormalized $NN$ potential $V_{\rm low-k}$ preserves the
observables predicted by $V_{NN}$ for the two-nucleon system,
and consequently its $\chi^2/{datum}$, up to the cutoff momentum
$\Lambda$. 
The $V_{\rm low-k}$ is a smooth potential and can be used directly in 
nuclear structure calculations within a perturbative approach.

While the $V_{\rm low-k}$ is defined in the momentum space, we perform
our calculations for finite nuclei in the coordinate space employing
a truncated harmonic oscillator (HO) basis.
Since the $V_{\rm low-k}$ procedure decouples the momentum space into
the low- and high-momentum regime, it is desirable to recover such a 
decoupling in the HO space.

Let us consider the relative motion of two nucleons in a harmonic
oscillator well in the momentum representation.
For a given maximum relative momentum $\Lambda$, the corresponding
maximum value of the energy is:

\begin{equation}
E_{\rm max} = \frac{ \hbar^2 \Lambda^2}{M}~~,
\end{equation}

\noindent
where $M$ is the nucleon mass.

This relation may be re-written in terms of the maximum number 
$N_{\rm max}$ of HO quanta for the relative coordinate system, so for
a given HO parameter $\hbar \omega$ we have:

\begin{equation}
\left( N_{\rm max} + \frac{3}{2} \right) \hbar \omega = \frac{ 
\hbar^2 \Lambda^2}{M}~~.
\end{equation}

The above equation provides a simple criterion to map out the
two-nucleon HO model space.
If we write the two-nucleon states as the product of HO wave functions

\begin{equation}
|a~b \rangle = | n_a l_a j_a,~n_b l_b j_b \rangle~~,
\end{equation}

\noindent
our HO model space is defined as spanned by those two-nucleon states that
satisfy the constraint

\begin{equation}
2n_a+l_a+2n_b+l_b \leq N_{\rm max}~~.
\end{equation}

The need for such a boundary condition for our model space may be also 
pointed out by the following considerations.  
The momentum contents of the two-nucleon wave function clearly depend
on the choice of the model space. 
For example, the average momentum of HO wave functions is proportional 
to $(\hbar \omega)^{1/2}$.
Suppose we are doing a calculation using a small model space together
with a small $\hbar \omega$, so that the momentum contents of the
basis wave functions are practically all below 1.5 $\rm fm^{-1}$. 
In this case, we clearly should use this value for $\Lambda$. 
Let us consider another situation where the model-space wave functions 
have important momentum components of, say, up to 3.0 $\rm fm ^{-1}$. 
Then in this case, we need to use a larger $\Lambda$.

\begin{figure}[H]
\includegraphics[scale=0.6,angle=0]{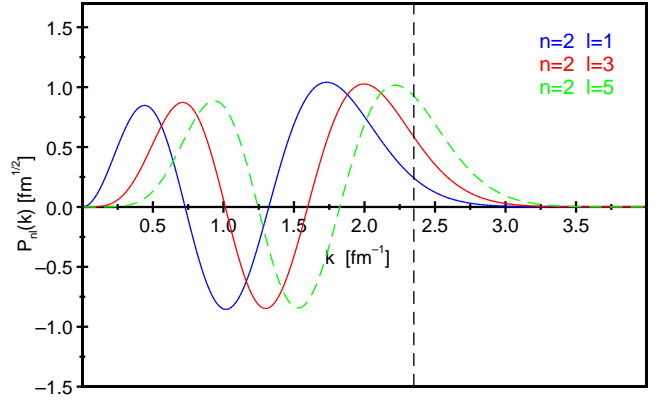}
\caption{Different $P_{nl}$'s for $\hbar \omega$=27 MeV. See text 
for details.}
\end{figure}

The above considerations may be illustrated by inspecting the explicit 
expression of the $V_{\rm low-k}$ matrix elements in terms of wave
functions of the center-of-mass and relative coordinates:

\begin{eqnarray}
\langle n_a l_a j_a, n_b l_b j_b ; J T | V_{\rm low-k} | n_c l_c j_c, n_d
l_d j_d; J T \rangle =  \sum_{LL'S} (-1)^{L+L'} \nonumber \\ 
\hat{j_a} \hat{j_b} \hat{j_c} \hat{j_d} (2L+1) (2L'+1)(2S+1) 
\left\{ \begin{array}{ccc}
l_a~ \frac{1}{2}~  j_a  \\
l_b~ \frac{1}{2}~  j_b  \\
 L~  S~  J  \end{array} \right\}
 \left\{ \begin{array}{ccc}
l_c~ \frac{1}{2}~  j_c  \\
l_d~ \frac{1}{2}~  j_d  \\
 L'~  S~  J  \end{array} \right\}  \nonumber \\
\sum_{n l n' l' N {\mathcal L}} [ 1-(-1)^{l+S+T} ]
\langle n l N {\mathcal L},~L | n_a l_a n_b l_b,~L \rangle~~~~~~~~~~~~~~~~~~~ \nonumber \\
\langle n' l' N {\mathcal L},~L' | n_c l_c n_d l_d,~L' \rangle 
\sum_{{\mathcal J}} (2{\mathcal J}+1) 
\left\{ \begin{array}{ccc}
{\mathcal L}~ l~  L  \nonumber \\
 S~  J~  {\mathcal J}  \end{array} \right\}
\left\{ \begin{array}{ccc}
{\mathcal L}~ l'~  L'  \\
 S~  J~  {\mathcal J}  \end{array} \right\} \\
\langle n l S T {\mathcal J} | V_{\rm low-k} | n' l' S T {\mathcal J}
\rangle ~~, ~~~~~~~~~~~~~~~~~~~~~~~~~~~~~~~~~~~~~~~~~~~~
\end{eqnarray}

\noindent
where $\hat{j}=\sqrt{2j+1}$, $\langle n l N {\mathcal L},~L | n_a
l_a n_b l_b,~L \rangle$ are the Brody-Moshinsky transformation brackets
\cite{brody60}, and, according to (4), $2n+l$ and $2n'+l'$ are both
$\leq N_{\rm max}$.
The matrix element $\langle n l S T {\mathcal J} | V_{\rm low-k} 
| n' l' S T {\mathcal J} \rangle$ is expressed in terms of the 
momentum-space HO wave functions $P_{nl}$'s as

\begin{eqnarray}
\langle n l S T {\mathcal J} | V_{\rm low-k} | n'
l' S T {\mathcal J} \rangle = \int_{0}^{\Lambda} \int_{0}^{\Lambda}
dk~dk'~kk'~P_{nl}(k) \nonumber \\
P_{n'l'}(k') V_{\rm low-k}^{ll'ST{\mathcal J}} (k,k')~~.~~~~~~~~~~~~~~~~~~~~~~~~~~~~~~~~~~~~~~~~~
\end{eqnarray}

Because in Eq. (6) the integrals are evaluated up to $\Lambda$, it is 
desirable to throw away those $P_{nl}$'s which extend significantly 
above the cutoff momentum, and pertain therefore to the high-momentum 
regime.
We have verified that applying the constraint (4) amounts to neglect
all $P_{nl}$'s which extend their tail more than $\simeq 5 \%$ above 
$\Lambda$.
Fig. 1 is an explanatory picture where we plot, for a given $\hbar
\omega=$27 MeV, three momentum space HO wave functions $P_{nl}$'s
with $2n+l=$5 , 7, and 9, respectively, as functions of $k$.
The vertical dashed line denotes a value of $\Lambda=2.35 \; {\rm 
fm}^{-1}$, corresponding to $N_{\rm max}$=7.

Relation (4) has a general validity, it should be applied every
time the $V_{\rm low-k}$ matrix elements are calculated in the HO
basis.

In this paper, making use of the present approach, we have 
studied the ground state properties of doubly closed-shell nuclei
within the framework of the Goldstone expansion \cite{goldstone57}.
More explicitly, starting from the purely intrinsic hamiltonian 

\begin{equation}
H= \left( 1 - \frac{1}{A} \right) \sum_{i=1}^{A} \frac{p_i^2}{2M} + \sum_{i<j}
\left( V_{ij} - \frac{ {\rm {\bf p}}_i \cdot {\rm {\bf p}}_j }{MA} \right)~~, 
\end{equation}

\noindent
where $V_{ij}$ stands for the renormalized $V_{NN}$ potential plus the 
Coulomb force, we construct the Hartree-Fock (HF) basis expanding the
HF single particle (SP) states in terms of HO wave functions.
The HF basis is then employed to sum up the Goldstone expansion, including 
contributions up to fourth-order in the two-body interaction.

Our calculations are made in a truncated model space, whose size 
is related to the values of the cutoff momentum $\Lambda$ and the $\hbar 
\omega$ parameter.
The calculations are performed increasing the $N_{\rm max}$ value
(and consequently $\Lambda$) and varying the $\hbar \omega$ value
until the dependence on $N_{\rm max}$ ($\Lambda$) is minimized.

\begin{figure}[H]
\includegraphics[scale=0.5,angle=0]{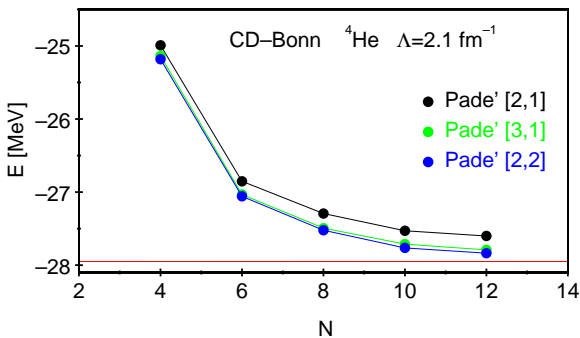}
\caption[ ]{$^4$He ground state energy as a function of the number of 
harmonic-oscillator major shells. The employed two-body interaction 
is a $V_{\rm low-k}$ derived from CD-Bonn potential with $\Lambda=2.1 
\; {\rm fm}^{-1}$. The oscillator parameter $\hbar \omega$ is equal to
18 MeV. Calculations have been performed with different Pad\'e
approximants, the red line representing an exact calculation.
\cite{viviani05b}.}
\end{figure}

\section{Convergence properties of the perturbative expansion}
In section IV we will show the calculated ground state energies of
some doubly closed-shell nuclei within the framework of the
perturbative approach (Goldstone expansion).
Here, we study the convergence properties of the perturbative series.
To this end, we have performed a test calculation, starting from a 
$V_{\rm low-k}$ with a fixed cutoff momentum $\Lambda=2.1 \; {\rm 
fm}^{-1}$ and derived from the $NN$ CD-Bonn potential \cite{cdbonn01}.
For this $V_{\rm low-k}$, hermitized according to the procedure based 
on the Cholesky decomposition suggested in Ref. \cite{andreozzi96}, an
exact calculation of the ground state energy of $^4$He has been
performed in the framework of the hyperspherical harmonic (HH)
approach \cite{viviani05}.
The value obtained is -27.95 MeV \cite{viviani05b}, considering the 
$V_{\rm low-k}$ as a $NN$ potential defined in the whole momentum
space, with $V_{\rm low-k}(k,k') =0$ when $k$ or $k' > \Lambda$.
With the above potential we have calculated the same quantity in the
framework of the Goldstone expansion.

Using Pad\'e approximants \cite{baker70,ayoub79} one may obtain a value
to which the perturbation series should converge.
We consider the following three Pad\'e approximants:

\begin{equation}
[ L|1 ]=E_0+E_1+...+\frac{E_L}{1-E_{L+1}/E_L}~~,
\end{equation}

\noindent
where $L=2,~3$, and

\begin{equation}
[ 2|2 ]=
\frac{E_0(1+\gamma_1+\gamma_2)+E_1(1+\gamma_2)+E_2}{1+\gamma_1+
\gamma_2}~~,
\end{equation}

where 

\[
\gamma_1 = \frac{E_2E_4-E_3^2}{E_1E_3-E_2^2} ~~,~~~~~
\gamma_2 = -\frac{E_3+E_1 \gamma_1}{E_2} \nonumber ~~,
\]

$E_i$ being the $i$th order energy contribution in the Goldstone
expansion.

In Fig. 2 the $^4$He calculated ground state energy is plotted versus 
the number of HO major shells included in the calculation, the red
line representing the exact value.
It is worth here to mention that no further truncation for the
two-nucleon states has been performed, we considering the $V_{\rm low-k}$
as defined in the whole momentum space.

\begin{figure}[H]
\includegraphics[scale=0.5,angle=0]{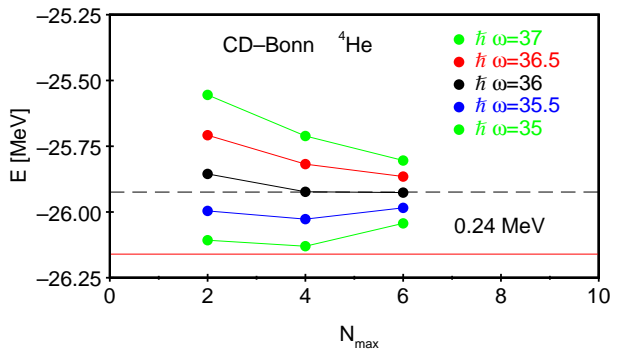}
\caption[ ]{$^4$He ground state energy with CD-Bonn potential as 
function of $N_{\rm max}$, for different values of $\hbar \omega$. 
The straight line represents the Faddeev-Yakubovsky result.}
\end{figure}

\begin{figure}[H]
\includegraphics[scale=0.5,angle=0]{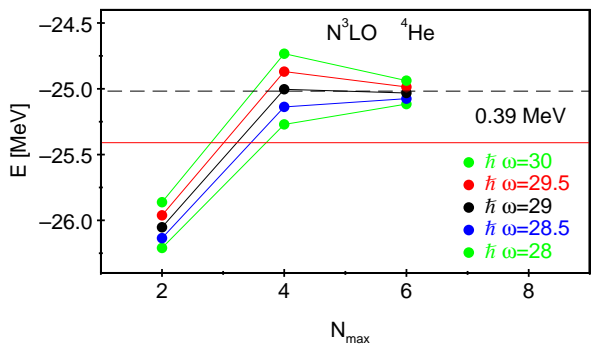}
\caption[ ]{Same as Fig.3, but for N$^3$LO potential.}
\end{figure}

We see that the Pad\'e approximants $[3|1]$ and $[2|2]$ give, for a
given number of HO major shells, almost the same value , the
difference being at most 45 keV.
The Pad\'e approximant $[2|1]$ differs at most by 200 keV from the 
$[3|1]$, and 245 keV from the $[2|2]$.
Moreover, the results, corresponding to the largest space we have
employed, come close to the exact value, the energies being -27.79 and 
-27.84 MeV with the $[3|1]$ and $[2|2]$ approximants, respectively.

On these grounds, we report in the following section the results obtained
using the Pad\'e approximant $[2|2]$.

\begin{figure}[H]
\includegraphics[scale=0.5,angle=0]{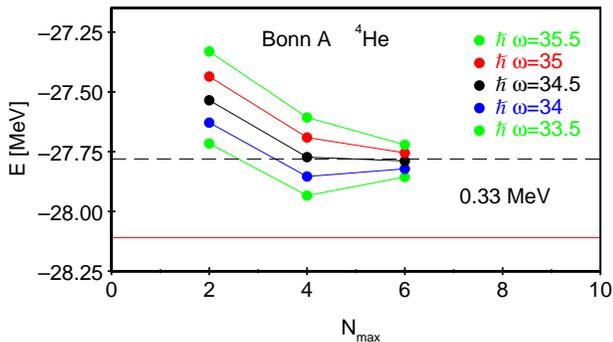}
\caption[ ]{Same as Fig.3, but for Bonn A potential.}
\end{figure}

\begin{figure}[H]
\includegraphics[scale=0.5,angle=0]{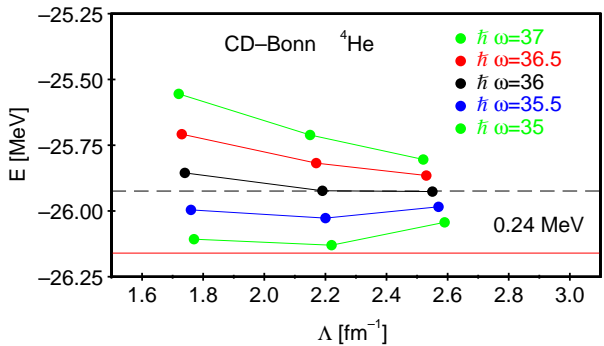}
\caption[ ]{$^4$He ground state energy with CD-Bonn 
potential as function of cutoff momentum $\Lambda$, for different
values of $\hbar \omega$. The straight line represents the
Faddeev-Yakubovsky result.}
\end{figure}

\section{Results}
To test the reliability of our calculations we have calculated the 
binding energy of $^4$He starting from different $V_{NN}$'s, and
compared our results with those obtained through the
Faddeev-Yakubovsky (FY) method.

In Figs. 3, 4, and 5 we show the calculated $^4$He ground state energies
obtained from the CD-Bonn \cite{cdbonn01}, N$^3$LO \cite{entem03}, and
Bonn A \cite{mach87} $NN$ potentials, respectively.
In each figure the straight red line indicates the FY result 
\cite{gloeckle93,nogga05} while the other curves represent our
calculated values, for different values of $\hbar \omega$, versus the
maximum number of HO quanta $N_{\rm max}$ that binds the two-nucleon 
configurations according to the relation (4).
In Fig. 6, we report the same results of Fig. 3, but versus the cutoff 
momentum $\Lambda$.

For the CD-Bonn, N$^3$LO, and Bonn A potentials we obtain convergence 
with $\hbar \omega=$ 36, 29, and 34.5 MeV, respectively.
Our calculated energies are -25.92, -25.02, and -27.78 MeV for
the above $V_{NN}$'s. 
These values are in good agreement with the FY results, the largest 
discrepancy being 0.39 MeV for N$^3$LO potential.
We have done similar calculations starting from other modern $NN$
potentials, such as the Nijmegen II \cite{stoks94} and Argonne V18 
\cite{wiringa95} potentials, but because of their stronger tensor 
components it has not been possible to achieve convergence.

\begin{figure}[H]
\includegraphics[scale=0.5,angle=0]{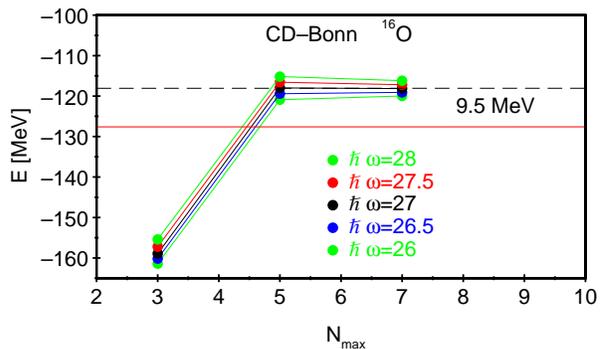}
\caption[ ]{$^{16}$O ground state energy with CD-Bonn potential as 
function of $N_{\rm max}$, for different values of $\hbar \omega$. 
The straight line represents the experimental value \cite{audi03}.}
\end{figure}

We have calculated also the ground state energies of $^{16}$O and
$^{40}$Ca starting from the CD-Bonn potential.
In Fig. 7 we present our results for $^{16}$O and compare them with
the experimental datum. The converged value, as can be seen from the
figure, is obtained for $\hbar \omega=$ 27 MeV and is equal to
-118.1 MeV, the discrepancy with the experimental value being 9.5 MeV.
In this case we cannot compare our calculations with the exact ones.
It is worth mentioning, however, the work by Fujii et al. 
\cite{fujii04} who, using the unitary model-operator approach (UMOA), 
predict for the CD-Bonn potential a converged value of -115.61
MeV, including only two-body correlations. 
In a more recent paper \cite{fujii05}, the above authors have
estimated the three-body cluster effect to contribute about -4 MeV.

As regards $^{40}$Ca, a calculation of its ground state energy
including fourth-order contributions in the Goldstone expansion has not
been done because of the large CPU time needed.
We therefore report in Fig. 8 the results obtained with the Pad\'e
approximant $[2|1]$, taking into account up to third-order contributions
in the Goldstone expansion.
The converged value is -307.8 MeV with $\hbar \omega=$ 25.5 MeV. 
In this case, the discrepancy with respect to the experimental value
is 34.2 MeV. 

\section{Summary}
In this work, we have calculated the ground state energy of some
doubly closed-shell nuclei in the framework of the Goldstone
expansion, starting from different realistic $NN$ potentials.
The short-range repulsion of these potentials has been renormalized by
integrating out their high-momentum components through the so-called
$V_{\rm low-k}$ approach.
We have introduced a criterion to map out the model space made up by
the two-nucleon states in the HO basis, according to the value of the 
cutoff momentum $\Lambda$.

\begin{figure}[H]
\includegraphics[scale=0.5,angle=0]{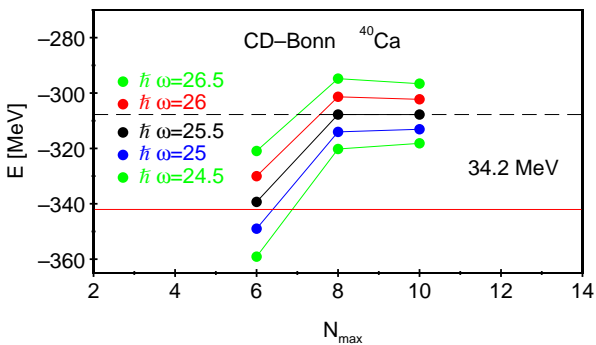}
\caption[ ]{Same as Fig.7, but for for $^{40}$Ca}
\end{figure}

The reliability of this procedure has been tested by calculating
the ground state energy of $^4$He, with the CD-Bonn, N$^3$LO, and Bonn A
potentials and comparing the results with the FY ones.
We have found that the energy differences are at most 390 keV. 
These differences are due to two reasons.
On the one hand, our calculations have been performed using a
perturbative approach, so that small contributions coming from higher
order terms may have not been completely taken into account by the
Pad\'e approximants.
On the other hand, we do not expect that Eq. (2) recovers exactly in 
the HO basis the $V_{\rm low-k}$ decoupling into low- and high-momentum
regime.

In any case, the limited size of the discrepancies shows that our approach 
provides a reliable way to renormalize the $NN$ potentials preserving
not only the two-body  but also the many-body physics.

On the above grounds, we have performed similar calculations for
heavier systems, such as $^{16}$O and $^{40}$Ca and obtained converged 
results using  model spaces not exceeding $N_{\rm max}=10$.

The rapid convergence of the results with the size of the HO model
space makes it very interesting to study in a near future heavier
systems employing our present approach.

\begin{acknowledgments}
We are indebted to Dr Michele Viviani for providing us with the
results of the calculation for $^4$He g.s. energy within the HH
approach. 
This work was supported in part by the Italian Ministero
dell'Istruzione, dell'Universit\`a e della Ricerca  (MIUR), and by the
U.S. DOE Grant No.~DE-FG02-88ER40388.
\end{acknowledgments}

\end{document}